\documentclass[twocolumn]{aastex631}

\begin{document}
\title{Photospheric Swirls in a Quiet-Sun Region}

\correspondingauthor{Jiajia Liu}
\email{jiajialiu@ustc.edu.cn}

\author[0009-0008-8972-2726]{Quan Xie}
\affiliation{Deep Space Exploration Laboratory/School of Earth and Space Sciences, University of Science and Technology of China, Hefei 230026, PR China}
\affiliation{CAS Key Laboratory of Geospace Environment, Department of Geophysics and Planetary Sciences, University of Science and Technology of China, Hefei 230026, PR China}

\author[0000-0003-2569-1840]{Jiajia Liu}
\affiliation{Deep Space Exploration Laboratory/School of Earth and Space Sciences, University of Science and Technology of China, Hefei 230026, PR China}
\affiliation{CAS Key Laboratory of Geospace Environment, Department of Geophysics and Planetary Sciences, University of Science and Technology of China, Hefei 230026, PR China}

\author[0000-0003-1400-8356]{Chris J. Nelson}
\affiliation{European Space Agency (ESA), European Space Research and Technology Centre (ESTEC), Keplerlaan 1, 2201 AZ Noordwijk, The Netherlands}

\author[0000-0003-3439-4127]{
Robert Erd\'elyi}
\affiliation{Solar Physics and Space Plasma Research Centre (SP2RC),
School of Mathematics and Statistics, University of Sheffield, Sheffield S3 7RH, UK}
\affiliation{Department of Astronomy, Eötvös Loránd University, Budapest, Pázmány P. sétány 1/A, H-1117, Hungary}
\affiliation
{Gyula Bay Zoltan Solar Observatory (GSO), Hungarian Solar Physics Foundation (HSPF) Pet\H{o}fi t\'er 3., Gyula H-5700, Hungary}

\author[0000-0002-8887-3919]{Yuming Wang}
\affiliation{Deep Space Exploration Laboratory/School of Earth and Space Sciences, University of Science and Technology of China, Hefei 230026, PR China}
\affiliation{CAS Key Laboratory of Geospace Environment, Department of Geophysics and Planetary Sciences, University of Science and Technology of China, Hefei 230026, PR China}




\begin{abstract}
 Swirl-shaped flow structures have been observed throughout the solar atmosphere, in both emission and absorption, at different altitudes and locations, and are believed to be associated with magnetic structures. However, the distribution patterns of such swirls, especially their spatial positions, remain unclear. Using the Automated Swirl Detection Algorithm (ASDA), we identified swirls from the high-resolution photospheric observations, centered on Fe I 630.25 nm, of a quiet region near the Sun's central meridian by the Swedish $1$-m Solar Telescope. Via a detailed study of the locations of the detected small-scale swirls with an average radius of $\sim$300 km, we found that most of them are located in lanes between mesogranules (which have an average diameter of $\sim$5.4 Mm) instead of the commonly believed intergranular lanes. The squared rotation, expansion/contraction and vector speeds, and proxy kinetic energy are all found to follow Gaussian distributions. Their rotation speed, expansion/contraction speed, and circulation are positively correlated with their radius. All these results suggest that photospheric swirls at different scales and locations across the observational $56.5^{\prime \prime} \times 57.5^{\prime \prime}$ field-of-view (FOV) could share the same triggering mechanism at preferred spatial and energy scales. A comparison with our previous work suggests that the number of photospheric swirls is positively correlated with the number of local magnetic concentrations, stressing again the close relation between swirls and local magnetic concentrations: the number of swirls should positively correlate with the number and strength of local magnetic concentrations.
\end{abstract}






\section{Introduction} \label{sec:intro}

\begin{figure*}[ht!]
    \centering
    \includegraphics[width=1.0\textwidth]{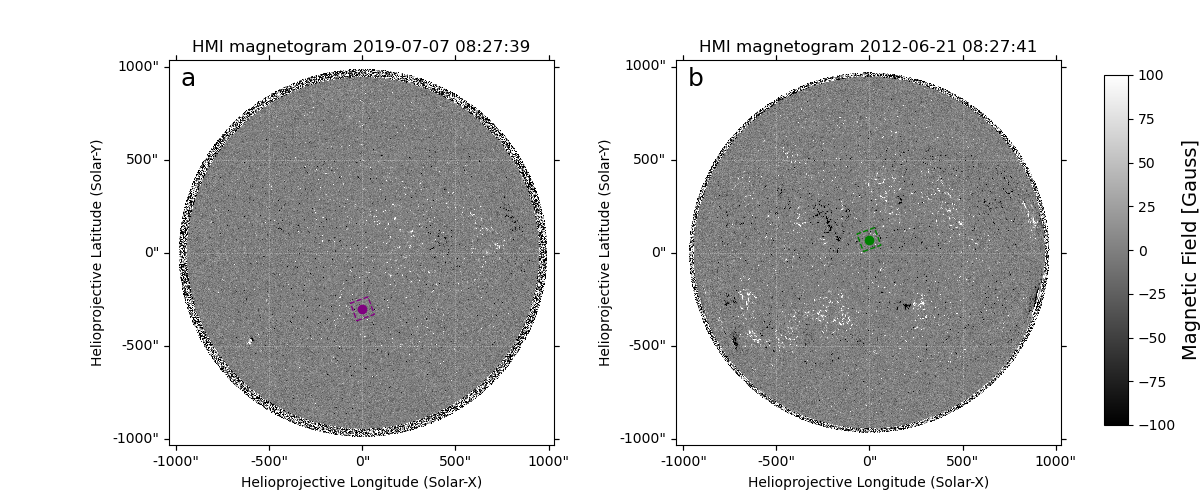}
    \caption{SDO/HMI photospheric magnetograms at 08:27:39 UT on 2019 July 07 (Left) and 08:27:41 UT on 2012 June 21 (Right), respectively. The purple and green rectangles represent FOVs of the SST observations in this study and \cite{liu2019automated}'s study.}
    \label{fig_fov}
\end{figure*}

Rotational phenomena, spanning from small-scale sub-arcsecond vortex flows \citep{bonet2008convectively} to large-scale solar tornadoes \citep[e.g.,][]{liu2012slow, li2012solar, su2012solar, wedemeyer2012magnetic, panesar2013solar, wang2016tornado}, have been extensively observed across various layers of the solar atmosphere \citep[e.g.,][]{liu2019automated, tziotziou2023vortex}. These dynamic motions are intricately linked to diverse mechanisms of energy transfer and conversion within the Sun, as well as other solar activities. For example, spicules (and dynamic fibrils) have also been observed to exhibit (possibly) rotational motion \citep[e.g.,][]{rompolt1975spectral, zaqarashvili2009oscillations, skogsrud2015temporal}. \cite{pike1998rotating} firstly described a structure known as magnetic tornadoes, which connects the convection zone to the upper solar atmosphere. \cite{wedemeyer2012magnetic} found that this structure can transport plasma to higher layers and \cite{li2012solar} observed a  larger solar tornado in a prominence and cavity. \cite{chmielewski2014numerical}
 found that, by numerical simulations, Alfv{\'e}n waves are associated with rotation of magnetic lines. Furthermore, \cite{liu2019evidence} provided evidence that photospheric swirls trigger Alfv{\'e}n pulses, which travel upwards into the upper chromosphere, carrying a considerable amount of energy and causing widespread chromospheric swirls.  Notably, small-scale swirls in the photosphere are believed to play a crucial role in energizing the upper solar atmosphere \citep{parker1983magnetic, velli1999alfven,shelyag2013alfven}. \cite{wedemeyer2009small} observed in Ca II 854.2 nm spectral line wing and wide-band images that groups of bright points in the photosphere exhibit relative motion at locations coinciding with chromospheric swirls. Furthermore, numerical simulations have suggested that swirls are associated with upward-directed Poynting flux in coronal loops, albeit diminishing with altitude \citep[e.g.,][]{shelyag2012vortices, snow2018magnetic, breu2023swirls}  and play a crucial role in transferring energy to the upper atmosphere \citep[e.g.,][]{shelyag2011photospheric, wedemeyer2012magnetic, murawski2018magnetic, yadav2020simulations, yadav2021vortex}. Some theoretical studies also have shown evidence that rotational motion will result in upward mass/momentum transfer \citep[e.g.,][]{scalisi2021propagation, scalisi2023generation}. 

Due to their importance in transferring energy in the solar chromosphere and their complex appearance, automated swirl identification is crucial in studying their statistics while minimizing human biases. The conventional methodology involves  estimating velocity fields from raw observational images and subsequently utilizing these fields to identify swirls. \cite{welsch2004ilct} and \cite{fisher2008subsurface} proposed the Fourier Local Correlation Tracking (FLCT) method for estimating velocity fields, while \cite{ramos2017deepvel} introduced a deep learning-based approach for the same purpose. Expanding upon these methodologies, \cite{liu2019automated} developed the Automated Swirl Detection Algorithm (ASDA) to automate swirl identification, leveraging algorithms introduced by \cite{graftieaux2001combining}. Similarly, \cite{dakanalis2021automated} devised an automated detection method for chromospheric swirls based on their morphological characteristics, and \cite{cuissa2022innovative} introduced SWIRL, another innovative automated swirl identification approach. Although the method proposed by \cite{dakanalis2021automated} does not rely on the horizontal velocity field estimated by Local Correlation Tracking (LCT) techniques, it is highly constrained by the choice of the right physical properties for defining a swirl. The SWIRL algorithm identifies and clusters swirls based on three important swirl characteristics  (local density $\rho$, spacing threshold $\delta$ and the $\gamma$ criterion), which are detailedly illustrated in \cite{cuissa2022innovative}, accurately recognizing swirls of various scales and demonstrating strong robustness against noise and shear flows. However, due to its high dependency on parameter selection, difficulties would be encountered when applied to  observational data as the parameters cannot be accurately adjusted according to different datasets.

Nevertheless, advancements in swirl identification techniques have significantly enhanced our understanding of swirl characteristics by providing physical parameters of swirls with smaller scales and shorter lifetimes. For instance, \cite{wedemeyer2012magnetic} found that chromospheric swirls typically have a diameter of approximately  1500 km. Estimates of chromospheric swirl density vary, ranging from 2 × 10$^{-3}$ Mm$^{-2}$ \citep{wedemeyer2012magnetic} to 8 × 10$^{
-2}$ Mm$^{-2}$ \citep{dakanalis2022chromospheric}, with observed lifetimes spanning from tens of seconds \citep{liu2019automated} to over 1.7 hours \citep{tziotziou2018persistent}.

Concerning photospheric swirls, \cite{balmaceda2010evidence} and \cite{vargas2011spatial} estimated the average radius of small-scale swirls to be around 1 Mm and 0.25 Mm, respectively. The occurrence rates for these swirls range from 1.4 × 10$^{-3}$ to 1.6 × 10$^{-3}$ swirls Mm$^{-2}$ min$^{-1}$. In contrast, large-scale photospheric swirls have diameters ranging from 1.5 to 21 Mm, with lifetimes exceeding 1 hour \citep{attie2009evidence}. More recent findings by \cite{liu2019automated} using images from the Solar Optical Telescope (SOT) on Hinode indicate approximately 1.62 × 10$^{5}$ swirls in the  entire photosphere at any instance of time, with an average radius of $\sim$290 km and an underestimated rotational speed below 1.0 km\ s$^{-1}$. These observations also reveal signatures of five-minute oscillation in both photospheric and chromospheric swirls, suggesting a potential role of the global $p$-mode in modulating them \citep{liu2023five}.

It is currently unclear whether there are differences in swirl distribution between different regions of the Sun, such as the northern and southern hemispheres, and active regions versus coronal holes or quiet regions. Different (or similar) distributions in different regions would indicate the different (or similar) formation mechanisms of swirls. In addressing the above questions, this paper takes the first step in a series of studies  by researching the overall distribution of thousands of detected small-scale swirls in a quiet region of solar photosphere and identifying their distribution characteristics. This current research is organised as follows: Section \ref{sec:data and method} outlines the methods for swirl identification used in this study. Section \ref{sec:results} presents a detailed and comprehensive analysis of swirl properties. Section \ref{diss and conclu} provides discussions and conclusions.


\section{Data and Method} \label{sec:data and method}
\begin{figure}[ht!]
    \centering
    \includegraphics[width=0.5\textwidth]{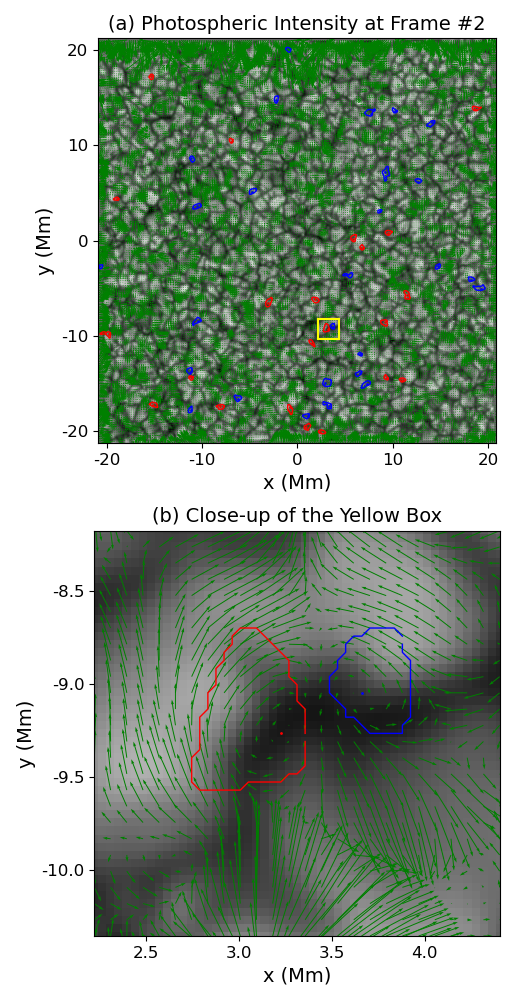}
    \caption{Detected swirls at frame 2. The background in panel (a) represents the photospheric intensity. Green arrows indicate the horizontal velocity field estimated by FLCT. Blue (red) dots and curves are the centers and edges of the detected swirls with counter-clockwise (clockwise) rotations. Panel (b) is the close-up view of the yellow box in panel (a).}
    \label{fig_swirl}
\end{figure}

\begin{figure*}
    \centering
    \includegraphics[width=1\textwidth]{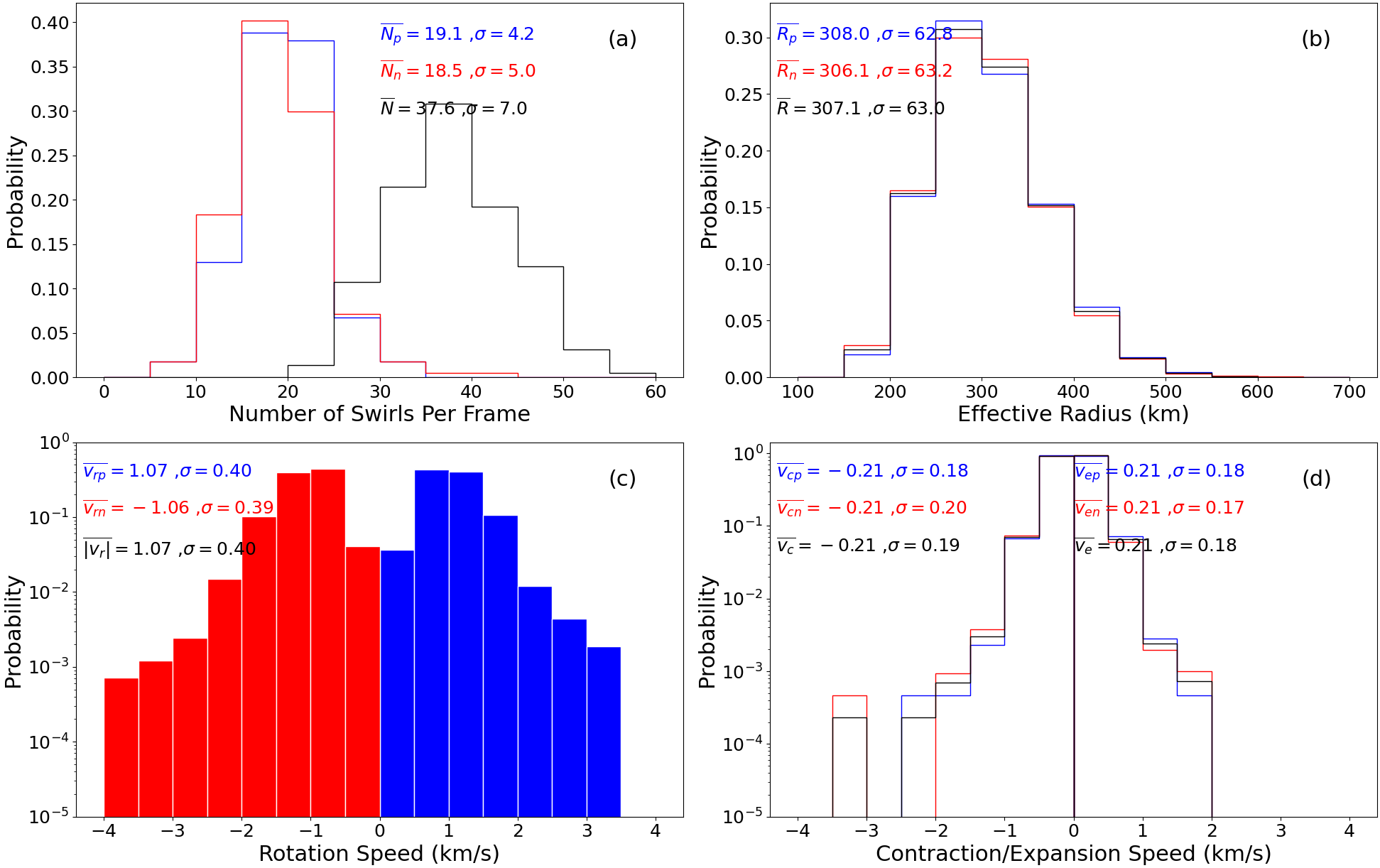}
    \caption{ Statistics of the number per frame ($N$), effective radius ($R$), rotation speed ($v_r$), and contraction/expansion ($v_c$/$v_e$) speed of all 8,424 photospheric swirls detected by ASDA from the SST Fe I 630.25 nm wide-band observation from 08:23:36 UT to 08:39:18 UT on 2019 July 7. Subscripts p and n (blue and red) denote positive and negative swirls, respectively. $\sigma$ in each frame is the corresponding standard deviation.}
    \label{fig_pn}
\end{figure*}

This study uses a dataset of photospheric images  centered on Fe I 630.25 nm,  with a spectral window width of 0.45 nm, obtained from the CRisp Imaging SpectroPolarimeter \citep[CRISP,][]{scharmer2006comments, scharmer2008crisp} on the Swedish $1$-m Solar Telescope \citep[SST,][]{scharmer20031}. The images were captured between 08:23:36 UT and 08:39:18 UT on July 7, 2019. The target region was a quiet-Sun area crossing the central meridian, centered at ($x_c=0^{\prime \prime}$, $y_c=-300^{\prime \prime}$), with a field of view (FOV) of $56.5^{\prime \prime} \times 57.5^{\prime \prime}$.  The pixel scale was $0.059^{\prime \prime}$($\sim$43.6 km), the spatial resolution was estimated to be at least 87.2 km which is twice the pixel size and the average cadence was 4.2 seconds. The FOV is tilted 70 degrees clockwise relative to the Sun's north pole (see Figure~\ref{fig_fov}a).  Meanwhile, an analog study conducted by \cite{liu2019automated} used data from the SST/CRISP instrument, collected between 08:07:22 UT and 09:05:44 UT on June 21, 2012, with a FOV of $55^{\prime \prime}\times55^{\prime \prime}$ (40.6 Mm × 40.6 Mm), centered at coordinates $x_c = - 3^{\prime \prime}$, $y_c=70^{\prime \prime}$ (see Fig~\ref{fig_fov}b). This data will be revisited in the discussions in Section \ref{diss and conclu}.

The Automated Swirl Detection Algorithm (ASDA) developed by \cite{liu2019automated} was used to detect photosphere swirls. The algorithm involves two key steps: (1) using Fourier Local Correlation Tracking (FLCT) to estimate the velocity field \citep[][]{welsch2004ilct, fisher2008subsurface}, and (2) applying ASDA to identify swirls within this field.  We set the pixel width of the Gaussian filter (sigma) to 10 and skip to None. The threshold and low-pass spatial filtering are left undefined with the bias correction (new feature in FLCT 1.06) turned on. This means we calculate the velocity for each pixel. More details of these parameter settings can be found in \cite{fisher2008subsurface}. It is worth mentioning that LCT has been proven to underestimated the velocity \citep{verma2013evaluating, liu2019co-spatial}. It is likely to influence the properties  (particularly the speeds, as already discussed in \cite{liu2019co-spatial}) of detected swirls. As FLCT has been used extensively in the past by the community and in our previous studies \citep[e.g.,][]{liu2019automated, liu2019evidence, liu2019co-spatial, liu2023five}, we also use it here for consistency. Several other methods of estimating the horizontal velocity field are currently tested as part of our another study.  The detection process relies on the metrics $\Gamma_1$ and $\Gamma_2$, as proposed by \cite{graftieaux2001combining}.  For each pixel \( P \), $\Gamma_1$(P) and $\Gamma_2$(P) are defined as:
\begin{equation}
    \begin{aligned}
        &\Gamma_{1}(P) =\boldsymbol{\hat{z}}\cdot\frac{1}{N}\sum_{S}\frac{\boldsymbol{n}_{PM}\times \boldsymbol{v}_{M}}{|\boldsymbol{v}_{M}|}, \\
        &\Gamma_{2}(P) =\boldsymbol{\hat{z}}\cdot\frac{1}{N}\sum_{S}\frac{\boldsymbol{n}_{PM}\times(\boldsymbol{v}_{M}-\overline{\boldsymbol{v}})}{|\boldsymbol{v}_{M}-\overline{\boldsymbol{v}}|}. 
    \end{aligned}
\end{equation}
Here, \( S \) represents a two-dimensional region with \( N \) pixels containing the target point \( P \). \( M \) is any point within region \( S \), \( \boldsymbol{\hat{z}} \) denotes the unit normal vector perpendicular to the observational surface pointing towards the observer, and \( \boldsymbol{n}_{PM} \) is the unit radius vector from point \( P \) to point \( M \). The variable \( \boldsymbol{\Bar{v}} \) represents the average velocity vector within region \( S \), while \( \boldsymbol{v}_M \) is the velocity vector at point \( M \). The symbols \( \times \) and \( \lvert \cdot \lvert \) denote the vector cross product and the magnitude (modulus) of vectors, respectively \citep{liu2019co-spatial}. The algorithm calculates $\Gamma_1$ and $\Gamma_2$ for each point in the velocity field, using 49 surrounding points as references. Points with $|\Gamma_1| \geq 0.89$ are identified as swirl centers, while those with $|\Gamma_2| \geq 2/\pi$ are defined as swirl boundaries. Positive (negative) values for $\Gamma_2$ indicate counterclockwise (clockwise) rotation, and the same holds for $\Gamma_1$. This approach allows us to identify swirl characteristics such as their locations, effective radii \citep[as defined by][]{liu2019automated}, rotation speeds, and expansion/contraction speeds. The underlying principles and methodology for obtaining these parameters are detailed in \cite{liu2019automated}.

\begin{figure}[ht!]
    \centering
    \includegraphics[width=0.5\textwidth]{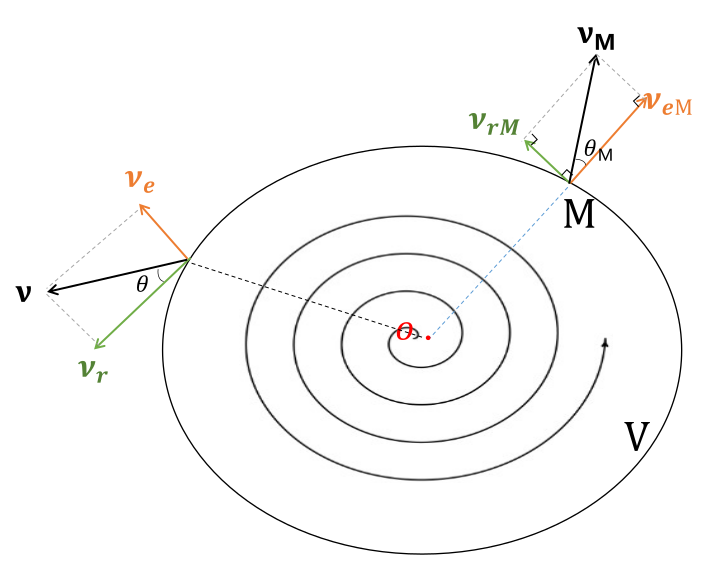}
    \caption{$S$ represents a swirl with the center $O$, and $v$ (black arrow) denotes its speed. $v_r$ (green arrow) and $v_e$ (orange arrow) are the results of the orthogonal decomposition of $v$, representing the rotation speed and expansion speed of the swirl, respectively.}
    \label{fig_vector}
\end{figure}
 Figure \ref{fig_swirl}(a) presents an example of the photospheric density distribution (in grayscale), with the velocity field estimated by FLCT overlaid as green arrows. The blue and red curves mark the locations and boundaries of detected swirls. At frame 2, 27 positive and 22 negative swirls are identified within the 41.8 $\times$ 42.5 Mm$^{2}$ field of view (FOV). Figure \ref{fig_swirl}(b) provides a close-up of the yellow box from panel (a), highlighting the edges and centers of two swirls — one rotating clockwise and the other counterclockwise.

\section{Results} \label{sec:results}
\subsection{Overall characteristics} \label{overall characteristics}
From the 224 velocity maps derived from 225 Fe I 630.25 nm wide-band images, a total of 8,880 swirls were detected.  But one swirl may exist in more than one frames, which means 8,880 is not the exact number of all detected swirls. So, to omit these repetitive swirls, firstly, method proposed by \cite{liu2019automated} is used to estimated lifetime of each swirl. Suppose two swirls, \( S_1 \) and \( S_2 \), are detected in two successive frames. \( S_1 \) is observed at time \( t_0 \) and \( S_2 \) at time \( t_0 + \Delta t \), where \( \Delta t \) is the observational cadence. \( S_1 \) and \( S_2 \) are considered the same swirl if the condition
\begin{equation}
    c_1 + v_{c_1} \cdot \Delta t \in S_2
\end{equation}
is satisfied. Here, \( c_1 \) is the center coordinate of \( S_1 \), and \( v_{c_1} \) is the velocity of its center. The symbol \( \in \) indicates that the predicted position belongs to swirl \( S_2 \).
Since a swirl’s rotational motion may change over time, we account for potential gaps in detection. Specifically, \( S_1 \) and \( S_3 \) (a swirl detected at \( t_0 + 2\Delta t \)) are still considered the same swirl if
\begin{equation}
    c_1 + v_{c_1} \cdot 2\Delta t \in S_3.
\end{equation}
\cite{liu2019automated} noted that the lifetime estimation of swirls that appear in only one frame are not fully reliable. So we also omit them when estimating the average lifetime which is then found to be 11.9 s.

Using this method, we identified which swirls were duplicates and removed them. After omitting the duplicates, we detected a total of 8,424 swirls, with 4,279 (50.8\%) rotating counterclockwise and 4,145 (49.2\%) rotating clockwise. The corresponding p-value is 0.14 ($>$0.05), which means that the number of swirls rotating counterclockwise is not significantly larger than thos rotating clockwise. The average swirl density in each frame is $2.11 \times 10^{-2}$  Mm$^{-2}$, within a field of view of 41.8 Mm $\times$ 42.5 Mm. 

\begin{figure*}[ht]
    \centering
    \includegraphics[width=1.0\textwidth]{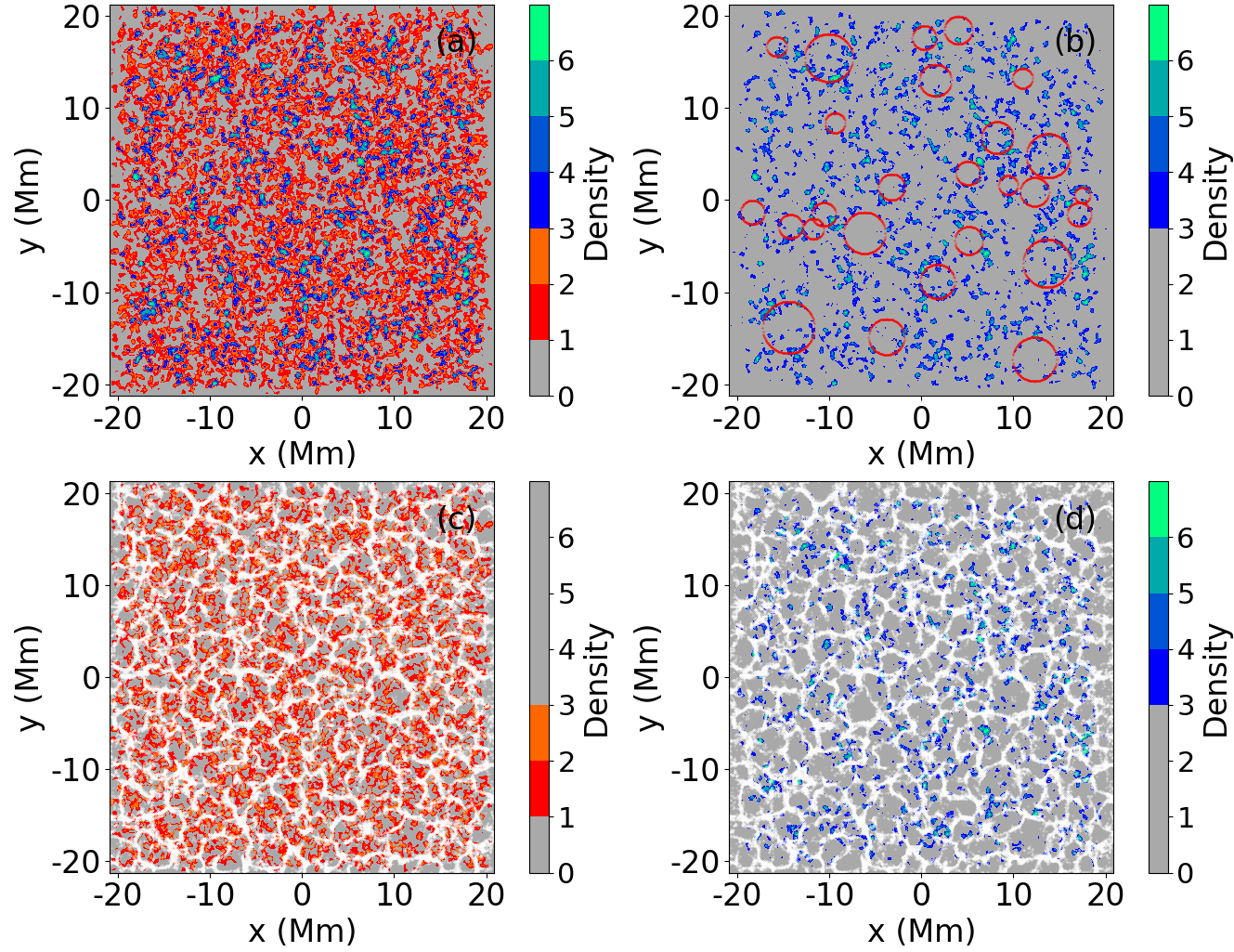}
    \caption{ Colors in (a) depict the number density of all 8,424 swirls. (b) is similar to panel (a) but only draws the region where density $\geq$ 3, other pixels are all set darkgray. Red circles indicate the location and size of vacancies encircled by the detected swirls, representing mesogranules (see text for details). (c) only depicts locations where density are 1 or 2. White dots (cork density $\rho_{cork}$ $>$ 2) outline the structure of mesogranules throughout the observational period. (d) is similar to panel (c) but only depicts locations where density are greater than or equal 3.}
    \label{fig_density}
\end{figure*}
The overall distribution characteristics of all 8,424 detected swirls are shown in Figure \ref{fig_pn}. Here, $N$ represents the number of swirls per frame, $R$ represents the swirl effective radius, $v_r$ represents the rotation speed, and $v_e$ and $v_c$ represent the expansion and contraction speeds, respectively. The speed of a swirl can be orthogonally decomposed into a rotation speed and an expansion/contraction speed. Here, $v_r$ represents the rotation speed of the swirl structure relative to its center, while $v_e$/$v_c$ represents the swirl's outward expansion/inward contraction speed (see Fig.~\ref{fig_vector} for an illustration). In this paper, speed $v$, rotation speed $v_r$, and expansion/contraction speed $v_e/v_c$ are velocity scalars, thus only representing values. The subscript $p$ denotes the characteristics of positive swirls (counterclockwise rotation), while the subscript $n$ denotes the characteristics of negative swirls (clockwise rotation). Positive swirls are distinguished by blue, and negative swirls by red. When no subscript is used, it represents all swirls in a given frame, depicted in black.

Figure \ref{fig_pn} shows that the number, effective radius, rotation speed, and expansion/contraction speeds of positive and negative swirls are nearly indistinguishable.  In each frame, the number of positive swirls slightly exceeds that of negative swirls (19.1 vs. 18.5). This difference may be attributed to the Coriolis force, given that the field of view is located in the southern hemisphere, or it may have occurred by chance. More statistical analysis on different regions at different latitudes are needed to confirm this. The average radius of these swirls is around 307 km, with an average rotation speed of approximately 1.07 km\ s$^{-1}$. About half of the swirls experience expanding or contracting, with the absolute values of both expansion and contraction speeds being approximately 0.21 km\ s$^{-1}$.

Therefore, we can conclude that there is no known difference in the studied distribution characteristics of positive and negative swirls throughout the entire observation period within the FOV. However, do their spatial distribution and temporal evolution also show no difference? This question will be explored and discussed in the next subsection.

\subsection{Spatial distribution} \label{spatial distribution}

A (number) density map was created to visually represent the distribution of these  8,424 swirls (Figure \ref{fig_density}a). Different colors at each pixel represent how many times it is located within a detected swirl from the first to the last frame in the observation.  The maximum density is 12, excluding 0, with the majority of values ranging between 0 and 6. As a result, we normalized the small fraction (0.68\%) of values greater than or equal to 7 by capping them at 6. Meanwhile, the density maps of positive and negative swirls generally exhibit a pattern of evenly interspersed distribution.  Figure \ref{fig_density}(b) shows a section of the density map from panel (a), highlighting only areas where the densities are not less than 3. These regions appear to form structures resembling larger ``granules". The red circles, chosen by eye, mostly encompass regions of high density in Figure \ref{fig_density}(b), indicating the locations and sizes of some of these larger ``granules" with varying diameters. On average, these larger ``granules" have a diameter of 5.2 \texttt{$\pm$} 1.1 Mm.

\begin{figure}[htbp]
    \centering
    \includegraphics[width=0.5\textwidth]{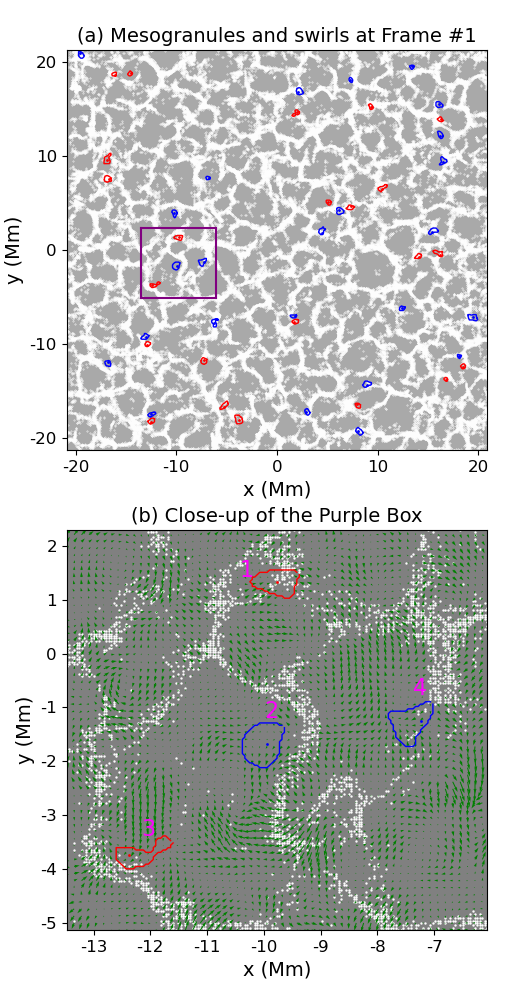}
    \caption{ (a) White dots where $\rho_{cork}$ $>$ 2 outline the inter-meosgranular lanes. Blue (red) dots and curves are the centers and edges of the detected swirls with counter-clockwise (clockwise) rotations at frame 1. (b) Close-up view of the purple box in panel (a). Green arrows indicate the horizontal velocity field estimated by FLCT. Numbers in pink are the sequential numbers of the detected swirls in the purple box.}
    \label{fig_meso}
\end{figure}

These larger ``granules'' are significantly smaller than supergranules whose average scale is $\sim$ 30 Mm \citep{rieutord2010sun}, but comparable to mesogranules with an average scale of 5 $\sim$ 10 Mm \citep{november1981detection}. Observations in \cite{november1981detection} showed clear evidence of the presence of meosgranules in scale between granule and supergranule for the first time. However, their results strongly depend on the extent to which they have canceled out the effects of oscillations and granules when determining the persistent velocities. Therefore, we use Lagrange tracers \citep[namely ``corks'',][]{chaouche2011mesogranulation} to study mesogranules over the entire observation period. Based on the two-dimensional velocity field of each frame within the observation interval, determined by FLCT, the corks move throughout the observation period. Since the displacement of corks between consecutive frames is less than one pixel due to the small velocities determined by FLCT, we update the positions of the corks at every 50 frames. After 945 seconds of motion, some corks converge to the same locations. Here, we cite a cork density function, $\rho_{cork}$,  representing the number of corks in each pixel \citep{chaouche2011mesogranulation}.  Figure \ref{fig_density}(c) and (d) show pixels where $\rho_{cork}$ $>$ 2, indicating these points are located in the lanes between mesogranular cells. Thus, white lines in the panels outline the structure of mesogranules, which have an estimated diameter of 4.2 \texttt{$\pm$} 0.7 Mm on average. It is evident from Figure \ref{fig_density}(c) and (d) that the locations where these 8,424 swirls occur once or twice are predominantly within the mesogranules, while regions where swirls frequently occur (i.e., 3 times or above) are mainly distributed along these white lanes.

 To quantify this, we define a criterion $N_c$, which represents the number of corks contained within a swirl, to determine whether a swirl is located in the inter-mesogranular lanes. For example, if $N_c$ $>$ 2, the swirl is classified as being located in the inter-mesogranular lanes. Figure \ref{fig_meso}(a) shows the mesogranules and detected swirls at frame 1. Figure \ref{fig_meso}(b) provides a close-up of the purple box from panel (a), where four swirls surrounding the mesogranules are marked. Swirls Nr. 1, 3, and 4 are classified as being in the inter-mesogranular lanes, as their $N_c$ values are all greater than 2, whereas Nr. 2 is not, with $N_c < 2$. Selection of 2 may be arbitrary to some extent. Therefore, we use a series of different criteria numbers (2, 3, 4 and 5) of $N_c$ to calculate the ratio of swirls located in inter-mesogranular lanes. Results are shown in Table \ref{ratio}. Moreover, a randomized experiment is conducted in order to further verify the reliability of the results. We distribute the swirls detected from all 224 frames randomly throughout the FOV to calculate their ratio along the inter-mesogranular lanes, also with 2, 3, 4 and 5 as the criteria. This process is repeated for 10000 times and results are also presented in Table \ref{ratio}. Under all four criteria, the authentic ratios fall well above the error ranges of the random test ratios, indicating that our previous observation that swirls frequently occur at inter-mesogranular lanes are not due to chance. We also calculate the p-value which is very close to 0 ($<$ 0.05), indicating our results are reliable at a 95\% confidence level. From Table \ref{ratio}, one can also see that there are approximately 60\% $\sim$ 70\% swirls located in inter-mesogranular lanes. This suggests that photospheric swirls detected from this particular observation tend to appear at inter-mesogranular lanes instead of inter-granular lanes suggested before \citep[e.g.,][]{liu2019automated}.
\begin{table}[ht]
\centering
\begin{tabularx}{0.5\textwidth}{lcc}
\toprule
Criterion & Authentic Ratio & Randomized Ratio \\
& \% & \% \\
\midrule
$N_c$ $>$ 2 & 71.9 & 64.9 $\pm$ 0.5  \\
$N_c$ $>$ 3 & 68.2 & 60.7 $\pm$ 0.5 \\
$N_c$ $>$ 4 & 64.6 & 57.3 $\pm$ 0.5 \\
$N_c$ $>$ 5 & 61.7 & 54.4 $\pm$ 0.5 \\
\bottomrule
\end{tabularx}
\caption{ Authentic and randomized ratio of swirls located in inter-mesogranular lanes. $N_c$ is the criterion defined to determine whether a swirl is located in the inter-mesogranular lanes.}
\label{ratio}
\end{table}

\subsection{Velocity distribution} \label{velocity distribution}
In Subsection \ref{overall characteristics}, we discussed the overall distribution characteristics of all swirls. Velocity, including rotation speed and expansion/contraction speed, is crucial to understanding swirl dynamics, thereby aiding in exploring the mechanisms behind swirl formation and dissipation. This subsection will focus on the velocity distribution of the  8,424 swirls detected in our study.

\begin{figure*}[htbp]
    \centering
    \includegraphics[width=1.0\textwidth]{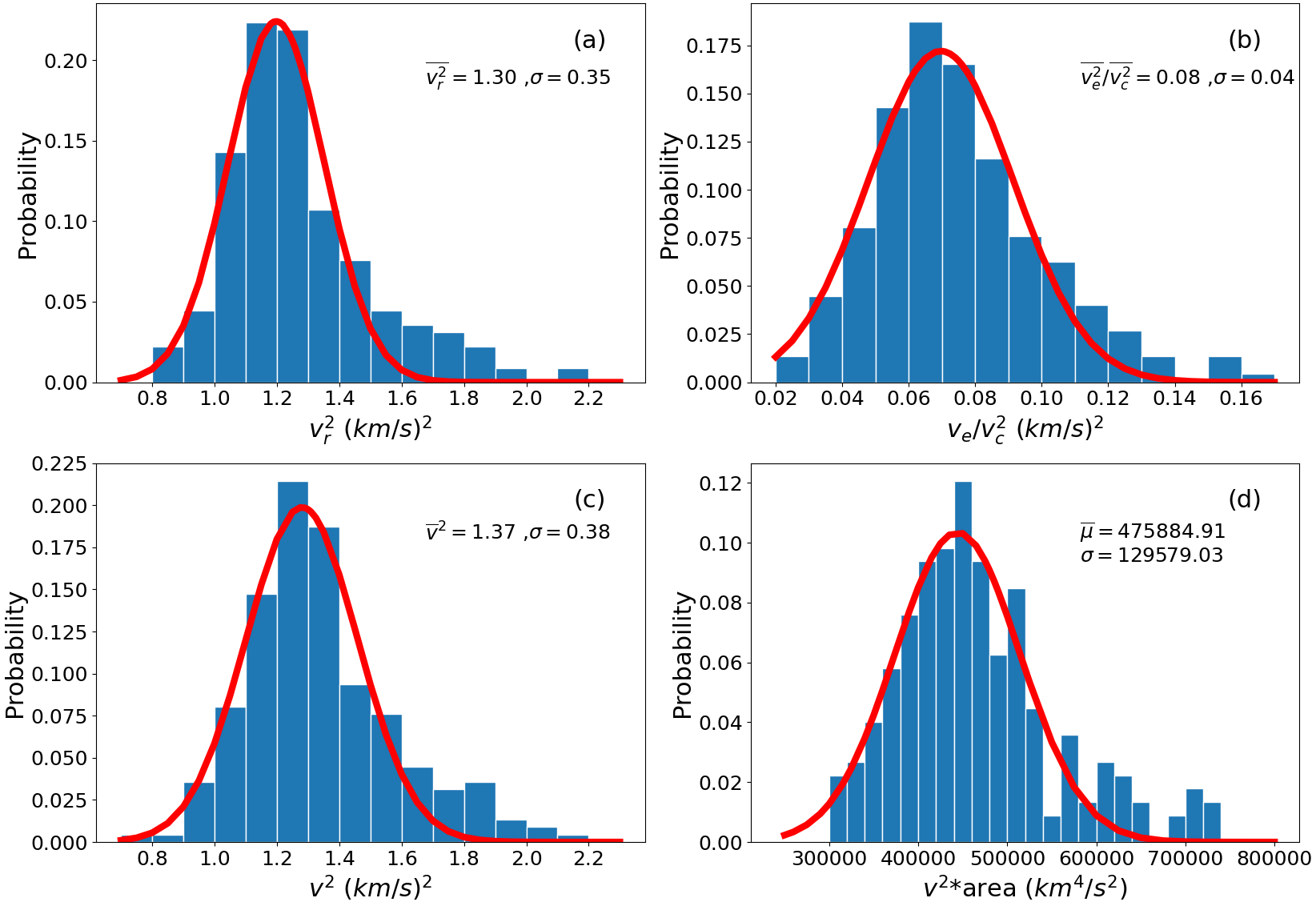}
    \caption{Statistics of the square of rotation speed ($v_r^2$), the square of contraction/expansion ($v_c^2$/$v_e^2$) speed, the square of speed ($v^2$) and the product of the square of speed ($v^2\cdot$area) and area of all 8880 photospheric swirls detected by ASDA from the SST Fe I 6302 Å wide-band observation from 08:23:36 UT to 08:39:18 UT on 2019 July 07. Red lines represent fitted Gaussian curves.}
    \label{fig_vel}
\end{figure*}

Figure \ref{fig_vel}(a)-(c) are the distributions of the square of the rotation speed ($v_r^2$), 
the square of expansion/contraction speed ($v_e^2$), and the squared speed norm ($v^2=v_r^2+v_e^2$) for all 8,424 
swirls, respectively. Figure \ref{fig_vel}(d) illustrates the distribution of the product of swirls' squared speed
norm and their area,  which we expect to be correlated with their kinetic energies, considering that most swirls are located in inter-mesogranular lanes and have similar observed intensities in the Fe I image. Red curves in the figures represent fitted Gaussian functions.  The cores of all four distributions in Figure~\ref{fig_vel} can be well modeled 
by Gaussian distributions, which however fail to capture the slow fall-off in the tails on the right side of each distribution. These Gaussian distributions suggest that the detected swirls are mostly excited at some particular energetic scale. Similar distributions were previously seen in swirls detected from both observations and realistic numerical simulations with the $p$-mode oscillation included \citep{liu2019co-spatial}, indicating that the preference 
of swirls in velocity and kinetic energy might be a result of the complex interaction between local motions and the 
global oscillation of the Sun \citep[e.g.][]{liu2023five}.

Let us explore the correlations among these properties with data on each swirl's speed  norm, rotation speed, and expansion/contraction speed. Focusing on larger swirls (radius greater than 500 km), \cite{brandt1988vortex} defined the swirl circulation along its boundary ($C$) as 
\begin{equation}\Gamma=\int_{C}\mathbf{v}  \cdot \mathrm{d
}\mathbf{l},\end{equation}
where $\mathbf{v}$ is the velocity, and $\mathrm{d}\mathbf{l}$ is the line element along curve $C$. 

\begin{figure*}[ht!]
    \centering
    \includegraphics[width=1.0\textwidth]{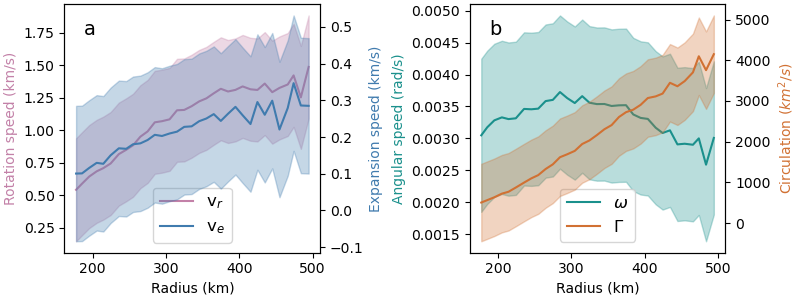}
    \caption{(a) Rotation speed ($v_r$) and expansion/contraction speed ($v_e/v_c$) as function of radius. (b) Angular speed ($\omega$) and circulation ($\Gamma$) as function of radius. The shaded areas in panel (a) and panel (b) respectively represent the error ranges for the curves of the corresponding colors.}
    \label{fig_rela}
\end{figure*}

Figure \ref{fig_rela}(a) shows the variation of rotation speed (purple curve) and expansion speed (blue curve) with respect to the radius, and Figure \ref{fig_rela}(b) depicts the angular speed (green curve) and circulation (orange curve) in relation to the radius. The shaded areas in both panels represent the error ranges of their respective curves.  These errors represent the variability in the data and are quantified as the standard deviations of the points along the curve, indicating the extent to which individual data points deviate from the curve's mean value. From panel (a), it is seen that the rotation and expansion/contraction speeds generally increase with radius. The fluctuations at the ends of the curve (for radii over 400 km) may be due to a decreasing number of events, with only 752 (8.5\%) swirls having a radius above 400 km. Meanwhile, both curves flattened when the radius is above $\sim$ 370 km.  It is worth further researching whether this is caused by the nature of swirls or by the inability of FLCT to estimate large speeds \citep{verma2013evaluating}. The quasi-linear increase of the rotation speed when the radius is less than $\sim$370 km suggests that the angular speed of these swirls might be close to a constant. Panel (b) indicates that the circulation also tends to increase with radius, consistent with what was found in \cite{brandt1988vortex}. The angular speed curve in panel (b) shows no clear trend. It varies within a narrow range (0.0026 $\sim$ {} 0.0036 rad\ s$^{-1}$), suggesting that swirls of different radii tend to have similar angular speeds and consistent with what was found from the relation between the rotation speed and the radius, as shown in panel (a).

\begin{figure*}[ht!]
    \centering
    \includegraphics[width=1\textwidth]{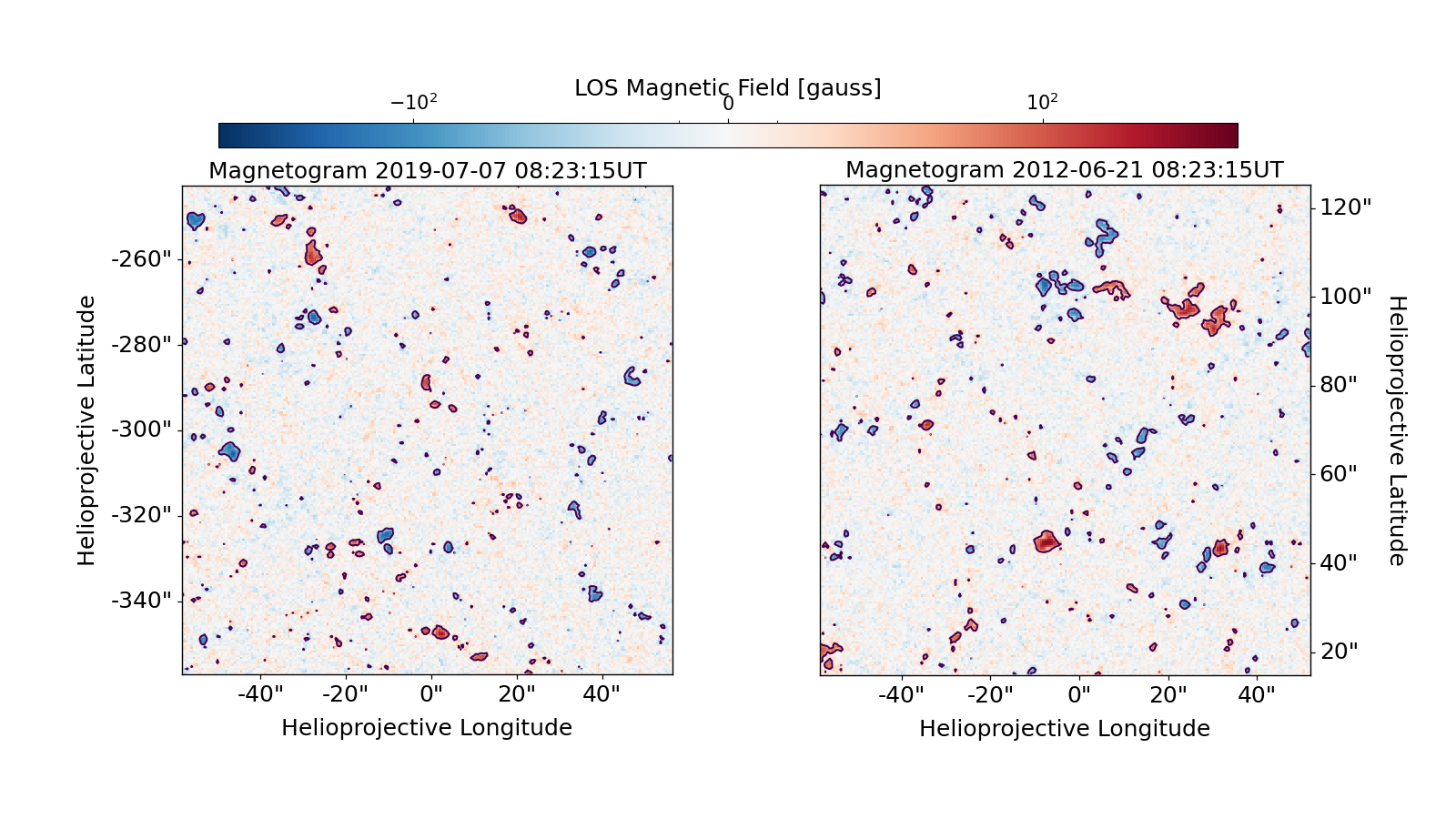}
    \caption{Magnetic field intensity diagrams at 08:23:15 on July 07, 2019, and June 21, 2012, with the black curve outlining an absolute magnetic field strength of 30 G.}
    \label{fig_hmi}
\end{figure*}
\section{Conclusions and discussions} \label{diss and conclu}

In this study, we used ASDA \citep{liu2019automated} and identified a total of  8,424 swirls in 224 frames of photospheric images taken at Fe \uppercase\expandafter{\romannumeral1} 630.25 nm by the Swedish $1$-m Solar Telescope (SST). The observed speed norm, rotation speed, expansion/contraction speed, and the  proxy kinetic energy of the detected swirls all follow Gaussian distributions.  The mean values of these Gaussian distributions indicate that photopsheric swirls are very likely excited at some particular spatial and energy scales, and the variability in swirls' properties about the mean values as shown by the Gaussian distributions are probably results of random processes. Regarding the spatial distribution of the  8,424 swirls, we found that, using Lagrange tracers, the identified swirls are  frequently concentrated along the inter-mesogranular lanes.

We also analyzed the relationships between swirl rotation speed, expansion/contraction speed, and circulation as a function of their radius, finding that all the above three parameters tend to increase as the radius increases. The circulation has a nearly linear relationship with radius, with a slope of about 2.72 km\ s$^{-1}$. The rotation speed and expansion/contraction speeds also show a nearly linear relationship within a radius of 350 km, with slopes of 4.0 $\times$ 10$^{-3}$\ s$^{-1}$ and 8.4 $\times$ 10$^{-4}$\ s$^{-1}$, respectively. These results, together with the almost invariant angular speed of swirls found in Figure~\ref{fig_rela}, again suggest that photospheric swirls are generated with the same underlying physical driver at some preferred spatial and energy scales. How these scales are related to the global and local flows of the Sun is an open question that needs more observations and theoretical studies or numerical simulations to explore.

 In Section \ref{sec:data and method}, it has been mentioned that \cite{liu2019automated} conducted a similar study using data observed in 2012, which is also collected from the SST/CRISP instrument (see Fig~\ref{fig_fov}b). Comparing the results from both studies, we find that the average radius, rotation speed, and expansion/contraction speed for positive and negative swirls are identical within errors. The only known difference is that the average number of positive swirls per frame (20.2) and negative swirls per frame (19.4) in our study is more than double that (9.1 and 9.2) in \cite{liu2019automated}, with the total number of swirls per frame (39.6) also more than double of their result (18.3). Given that our FOV (41.8 Mm × 42.5 Mm) is similar to the FOV in \cite{liu2019automated} (40.6 Mm × 40.6 Mm), it is clear that the swirl number density in our study is twice of that in \cite{liu2019automated}.

This finding is intriguing, especially considering that swirls were suggested to be closely related to local magnetic concentrations in realistic numerical simulations \citep{liu2019co-spatial}. An apparent difference between these two observations is that 2019 was near the solar minimum, while 2012 was closer to the maximum. Generally, the average solar magnetic field strength in 2012 should be higher than in 2019. However, the number of swirls in 2019 exceeds their counterpart in 2012. To investigate this difference, we analyzed the magnetic field strength distributions throughout the two observational periods, i.e., 08:07:22 UT to 09:05:44 UT on June 21, 2012, and 08:23:36 UT to 08:39:18 UT on July 7, 2019  using the line-of-sight magnetic field hmi.M\_45s, detected by Helioseismic Magnetic Imager \citep[HMI,][]{scherrer2012helioseismic} onboard the Solar Dynamics Observatory (SDO).

\begin{figure}[ht!]
    \centering
    \includegraphics[width=0.5\textwidth]{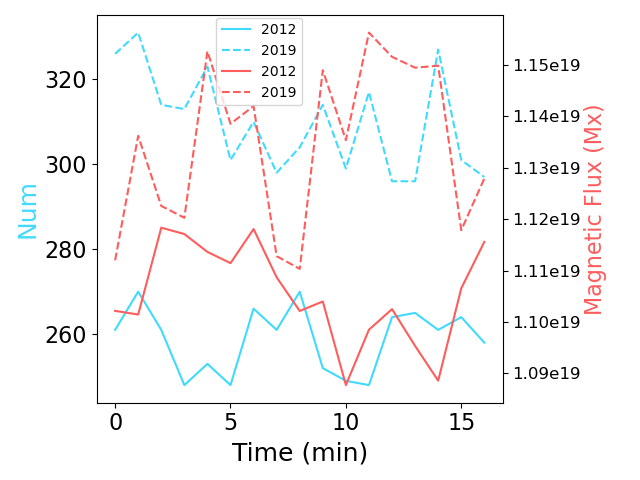}
    \caption{The solid and dashed lines represent data from 2012 and 2019, respectively. The sky blue curve represents the number of regions enclosed by the 30 Gauss contours, while the red curve represents the total absolute magnetic flux across all enclosed regions. The x-axis represents the time elapsed from the first frame of the observations.}
    \label{fig_compare}
\end{figure}

In Figure~\ref{fig_hmi}, red and blue colors depict local magnetic concentrations exceeding $\pm$ 30 G. A quick comparison between these two panels in Figure~\ref{fig_hmi} shows that, although the Sun was more active in 2012 than in 2019, there turns out to be more local magnetic concentrations (though smaller in size) in the FOV of the 2019 dataset.

Figure~\ref{fig_compare} shows a comparison between the number (sky blue curve) and total absolute magnetic flux (red curve) in the FOVs of the 2012 (solid curve) and 2019 (dashed curve) datasets and their evolution with time starting from the first frame of each observation. The 2019 dataset corresponds to a significantly higher number of magnetic concentrations and greater total absolute magnetic flux than the 2012 dataset. Specifically, the number of pixels with magnetic field strength greater than 30 G was 327 in 2019, compared to 261 in 2012. This, together with the fact that there were more swirls in the dataset of 2019 than in the counterpart dataset of 2012, indicates that the number of swirls should positively correlate with the number and strength of local magnetic concentrations. However, local magnetic concentrations have little impact on swirl radius, rotation speed, and expansion/contraction speed, given that the above swirl parameters derived from 2012 and 2019 datasets are similar. The above results might suggest that the number of swirls exhibits an anti-correlation with the solar cycle activity level. However, they could have also been caused by the different latitudes (and hemispheres) of the studied two datasets, especially considering that the magnetic field in the southern hemisphere was found generally larger than that of the northern hemisphere \citep[e.g.,][]{liu2023power}. To examine which of the above processes led to the observed different number densities of swirls in different quiet-Sun regions, a statistical study on a considerable number of high-resolution photospheric observations across an entire solar activity cycle is urgently needed.

\section*{Acknowledgements}
We acknowledge using Swedish $1$-m Solar Telescope (SST) data. SST is operated on the island of La Palma (Spain) by the Institute for Solar Physics of Stockholm University in the Spanish Observatorio del Roque de los Muchachos of the Instituto de Astrof\'isica de Canarias. J.L. and Q.X. acknowledge the support from the Strategic Priority Research Program of the Chinese Academy of Science (Grant No. XDB0560000), National Key Technologies Research, Development Program of the Ministry of Science and Technology of China (2022YFF0711402), and the National Natural Science Foundation (NSFC 12373056, 42188101). C.J.N. is thankful to ESA for support as an ESA Research Fellow. R.E. is grateful to the Science and Technology Facilities Council (STFC, grant No. ST/M000826/1) UK, acknowledges NKFIH (OTKA, grant No. K142987 and Excellence Grant TKP2021-NKTA-64) Hungary and PIFI (China, grant number No. 2024PVA0043) for enabling this research. This work was also supported by the International Space Science Institute project (ISSI-BJ ID 24-604) on "Small-scale eruptions in the Sun”.



\end{document}